\documentstyle[aps,epsf]{revtex}
\title{On the Evolution of Cosmological Type Ia Supernovae and the
Gravitational Constant}
\author{E. Garc\'\i a-Berro\footnote{garcia@fa.upc.es}}
\address{Departament de F\'{\i}sica Aplicada (IEEC/UPC), Universitat 
	 Polit\`ecnica de Catalunya, Jordi Girona Salgado s/n, M\`odul 
	 B--5, Campus Nord, 08034 Barcelona, Spain}
\author{E. Gazta\~naga\footnote{gaztanaga@ieec.fcr.es}, 
        J. Isern\footnote{isern@ieec.fcr.es}}  
\address{Institut d'Estudis Espacials de Catalunya (IEEC/CSIC), Edifici 
	 Nexus, Gran Capit\`a 2--4, 08034 Barcelona, Spain}
\author{O. Benvenuto\footnote{obenvenuto@fcaglp.fcaglp.unp.edu.ar}
 \& L. Althaus\footnote{althaus@fcaglp.fcaglp.unp.edu.ar}}
\address{Facultad de Ciencias Astron\'omicas y Geof\'\i sicas, Paseo
	 del Bosque s/n, (1900), La Plata, Argentina}

\date{\today}

\def\la{\mathrel{\mathpalette\fun <}}
\def\ga{\mathrel{\mathpalette\fun >}}
\def\fun#1#2{\lower3.6pt\vbox{\baselineskip0pt\lineskip.9pt
\ialign{$\mathsurround=0pt#1\hfill##\hfil$\crcr#2\crcr\sim\crcr}}}

\begin{document}

\maketitle

\begin{abstract}
There are at least three ways in which a varying gravitational 
constant $G$ could affect the interpretation of the recent 
high-redhisft Type Ia supernovae results. If the local value 
of $G$ at the space-time location of distant supernovae is 
different, it would change both the thermonuclear energy release 
and the time scale of the supernova outburst. In both cases 
the effect is related to a change in the Chandrasekhar mass 
$M_{\rm Ch}\propto G^{-3/2}$. Moreover the integrated variation 
of $G$ with time would also affect cosmic evolution and therefore 
the luminosity distance relation. Here we investigate in a consistent 
way how these different effects of a varying $G$ could change
the current interpretation of the Hubble diagram of Type Ia 
supernovae. We parametrize the variation of $G$ using scalar-tensor 
theories of gravity, such as the Jordan-Brans-Dicke theory or its 
extensions. It is remarkable that Dirac's hypothesis that $G$ 
should decrease with time can qualitatively explain the observed 
$\Delta m \simeq 0.2$ mag decrease at $z\simeq 0.5$ (with respect 
to a decelerating universe) and, at the same time, reduce the
duration of the risetimes of distant Type Ia supernovae as recently 
reported.
 \end{abstract}

\section{Introduction}

Type Ia supernovae (SNeIa) are supposed to be one of the best examples 
of standard candles. This is because, although the nature of their 
progenitors and the detailed mechanism of explosion are still the 
subject of a strong debate, their observational light curves are 
relatively well understood and, consequently, their individual 
intrinsic differences can be easily accounted for. Therefore, 
thermonuclear supernovae are well suited objects to study the 
Universe at large, especially at high redshifts $(z\sim 0.5)$, 
where the rest of standard candles fail in deriving reliable 
distances, thus providing an unique tool for determining 
cosmological parameters or discriminating among different
alternative cosmological theories.

Using the observations of 42 high redshift Type Ia supernovae and 18 
low redshift supernovae (Riess et al. 1998; Perlmutter et al. 1999),
both the Supernova Cosmology Project and the High-$z$ Supernova Search
Team found that the peak luminosities of distant supernovae appear 
to be $\sim 0.2$ magnitude fainter than predicted by a standard 
decelerating universe $(q_0>0)$. Based on this, the Supernova Cosmology 
Project derived $\Omega_{\rm M}=0.28^{+0.14}_{-0.12}$ at $1\sigma$, 
for a flat universe, thus forcing a non-vanishing cosmological 
constant. However this conclusion lies on the assumption that 
there is no mechanism likely to produce an evolution of the observed 
light curves over cosmological distances. In other words: both teams 
assumed that the intrinsic peak luminosity and the time scales of 
the light curve were exactly the same for both the low-$z$ and the 
high-$z$ supernovae. 

More recently Riess et al. (1999a,b) have found evidences of 
evolution between the samples of nearby supernovae and those 
observed at high redshifts by comparing their respective risetimes,
thus casting some doubts about the derived cosmological parameters.
In particular Riess et al. (1999a,b) find that the sample of low-$z$
supernovae has an average risetime of $19.98\pm 0.15$ days whereas the 
sample of high-$z$ supernovae has an average risetime of $17.50\pm
0.40$ days. The statistical likelihood that the two samples are 
different is high $(5.8\sigma)$. Riess et al. (1999b) also analyze 
several potential alternatives to produce, within a familiy of 
theoretical models, an evolution with the observed properties: 
distant supernovae should be intrinsically fainter and at the 
same time should have smaller risetimes. All the families of 
models studied so far have the inverse trend: decreasing peak
luminosities correspond to longer risetimes.

On the other hand, and from the theoretical point of view, it is
easy to show that a time variation of the gravitational constant, 
in the framework of a Scalar-Tensor cosmological theory, can 
reconcile the observational Hubble diagram of SNeIa with an open 
$\Omega_\Lambda=0$ universe.\footnote{While we were writing
this paper we became aware of a similar idea independently
proposed by Amendola et al. (1999).} The starting point is simple:
assume that all thermonuclear supernovae release the same amount
of energy $(E)$. In a simple model of light curve (Arnett 1982) the 
peak luminosity is proportional to the mass of nickel synthetized,
which in turn, to a good approximation, is a fixed fraction of the 
Chandrasekhar mass $(M_{\rm Ni}\propto M_{\rm Ch})$, which depends
on the value of gravitational constant: $M_{\rm Ch}\propto G^{-3/2}$.
Thus we have $E\propto G^{-3/2}$, and if one assumes a slow decrease
of $G$ with time, distant supernovae should be dimmer. Moreover,
the time scales of supernovae also depend on the Chandrasekhar 
mass. Let us elaborate on this last point. According to the analytic 
model of light curve of Arnett (1982), the width of the peak of the 
light curve of SNeIa is given by:

\begin{equation}
\tau\propto\Big(\frac{M_{\rm ej}^3}{M_{\rm inc}}\Big)^{1/4}
\label{risetime}
\end{equation}

\noindent
where $M_{\rm ej}$ is the ejected mass and $M_{\rm inc}$ is the 
incinerated mass. Within our current knowledge of the mechanisms
of explosion of SNeIa both masses can be considered proportional 
to the Chandrasekhar mass, and therefore we have $\tau\propto 
M_{\rm Ch}^{1/2}$ or, equivalently, $\tau\propto G^{-3/4}$. 
Since the risetime for distant supernovae is obtained from 
semi-empirical models, that is a template light curve which 
takes into account the decline rate and the width of the peak, one 
can then also assume this dependence on $G$ for the risetime. This 
expression has the right properties since distant supernovae have 
smaller peak luminosities and, at the sime time, smaller risetimes,
as required by observations.

\section{The effects of a varying $G$}

Despite the beauty and successes of the simplest version of General 
Relativity (GR), the possibility that $G$ could vary in space and/or
time is well motivated. Its study can shed new light into fundamental
physics and cosmology and it seems natural in Scalar-Tensor theories
of gravity (STTs) such as Jordan-Brans-Dicke (JBD) theory or its 
extensions.

To make quantitative predictions we will consider cosmic evolution
in STTs, where $G$ is derived from a scalar field $\phi$ which is 
characterized by a function $\omega=\omega(\phi)$ determining the 
strength of the coupling between the scalar field and gravity. 
In the simplest JBD models, $\omega$ is just a constant and $G 
\simeq \phi^{-1}$, however if $\omega$ varies then it can increase 
with cosmic time so that $\omega=\omega(z)$. The Hubble rate $H$ in 
these models is given by:
\begin{equation}
H^2 \equiv \left({{\dot a}\over{a}}\right)^2= {8 \pi \rho\over{3\phi}}
+{1\over{a^2 R^2}}+ {\Lambda\over{3}}+{\omega\over{6}}{{{\dot \phi}^2}
\over{\phi^2}} - H {{{\dot \phi}}\over{\phi}},
\label{H^2}
\end{equation}
this equation has to be complemented with the acceleration equations 
for $a$ and $\phi$, and with the equation of state for a perfect fluid: 
$p=(\gamma-1)\rho$ and ${\dot \rho}+3\gamma H\rho=0$. The structure of 
the solutions to this set of equations is quite rich and depends 
crucially on the coupling function $\omega(\phi)$ (see Barrow \& 
Parsons 1996). Here we are only interested in the matter dominated 
regime: $\gamma=1$. In the weak field limit and a flat universe the 
exact solution is given by:
\begin{equation}
G= {4+2\omega\over{3+2\omega}}\phi^{-1} = G_0 (1+z)^{1/(1+\omega)}.
\label{G(t)}
\end{equation}
In this case we also have that $a = (t/t_0)^{(2\omega+2)/(3\omega+4)}$. This 
solution for the flat universe is recovered in a general case
in the limit $t \rightarrow \infty$ and also arises as an exact 
solution of Newtonian gravity with a power law $G \propto t^{n}$ 
(Barrow 1996). For non-flat models, $a(t)$ is not a simple power-law 
and the solutions get far more complicated. To illustrate the effects
of a non-flat cosmology we will consider general solutions that 
can be parametrized as Eq.[\ref{G(t)}] but which are not simple
 power-laws in $a(t)$. In 
this case, it is easy to check that the new Hubble law given by 
Eq.[\ref{H^2}] becomes:
\begin{equation}
H^2(z) = H^2_0 ~\left[ ~\hat\Omega_M (1+z)^{3+1/(1+\omega)} +
\hat\Omega_R (1+z)^2 + \hat\Omega_\Lambda ~\right]
\label{H(z)}
\end{equation}
where $\hat\Omega_M$,$\hat\Omega_R$ and $\hat\Omega_\Lambda$ follow
the usual relation: $\hat\Omega_M+\hat\Omega_R+\hat\Omega_\Lambda=1$ 
and are related to the familiar local ratios ($z \rightarrow 0$): 
$\Omega_M \equiv 8\pi G_0 \rho_0/(3H_0^2)$, $\Omega_R=1/(a_0 R H_0)^2$ 
and $\Omega_\Lambda=\Lambda/(3H_0^2)$ by:
\begin{eqnarray}
\hat\Omega_M&=&\frac{\Omega_M}{g} 
~\left(\frac{4+2\omega}{3+2\omega}\right)   ~~~ ; ~~~
\hat\Omega_\Lambda=\frac{\Omega_\Lambda}{g} ~~~ ; ~~~ 
\hat\Omega_R=\frac{\Omega_R}{g}\\
g &\equiv& 1+\frac{1}{(1+\omega)}-\frac{1}{6}~\frac{\omega}
{(1+\omega)^2}
\label{omegahat}
\end{eqnarray}
Thus the GR limit is recovered as $\omega \rightarrow \infty$. The 
luminosity distance $d_L=d_L(z,\Omega_M,\Omega_\Lambda,\omega)$ is 
obtained as usual from the (line-of-sight) comoving coordinate distance:
$r(z)=\int dz'/H(z')$, with the trigonometric or the hyperbolic sinus
to account for curvature (Peebles 1993). In the limit of small $z$ we
recover the usual Hubble relation: $y= H_0 r= z-(1+\hat q_0) z^2/2$
where a new deceleration $\hat q_0$ parameter is related to the
standard one by:
\begin{equation}
\hat q_0 ~=~{q_0\over{g}} + {\hat\Omega_M\over{2(1+\omega)}}.
\end{equation}
One can see from this equation that even for relative small values
of $\omega$ the cosmological effect is small. For example for 
$\Omega_M\simeq 0.2$ and $\Omega_\Lambda \simeq 0.8$ we have $q_0 
\simeq -0.7$ while $\hat q_0$ is around $\hat q_0 \simeq -0.4$ 
for $\omega \simeq 1$. Note nevertheless that this effect, although
small, tends to decrease the acceleration and therefore it partially
decreases the effect in the peak luminosity of SNeIa caused by an 
increasing $G$. In summary, Eq.[\ref{G(t)}] parametrizes the change 
in $G$ as a function of $\omega$ while Eqs.[\ref{H(z)}-\ref{omegahat}] 
parametrize the corresponding cosmic evolution. 

\begin{figure}
\epsfxsize 5in
\epsfbox{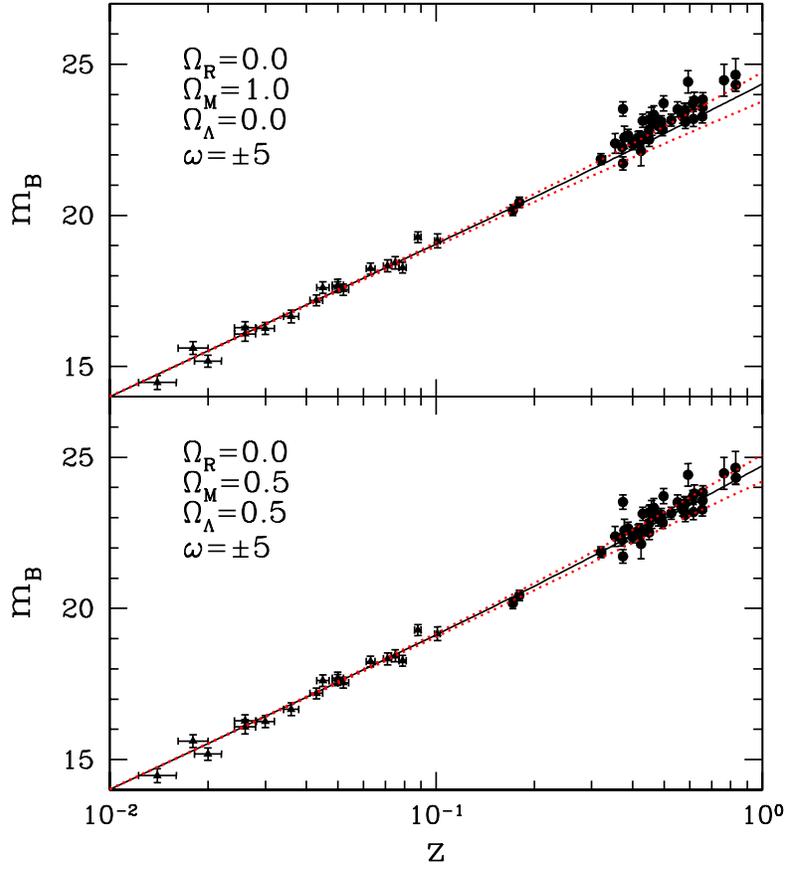}
\caption{Hubble diagram for the high-redshift SNe.}
\label{fig1}
\end{figure}

As mentioned in the introduction, we are assuming that thermonuclear 
supernovae release a similar amount of energy $E\propto G^{-3/2}$. 
Thus using Eq.[\ref{G(t)}], we have:
 
\begin{equation}
\frac{E}{E_0}=\big(\frac{G}{G_0}\big)^{-3/2}
~~~; ~~~ M-M_0=\frac{15}{4}\log\big(\frac{G}{G_0}\big)=
 \frac{15}{4(1+\omega)}\log\big(1+z\big),
\end{equation}
were $M$ is the absolute magnitude and the subscript 0 denotes the
local value. Therefore we have the following Hubble relation:
\begin{equation}
m(z)=M_0 +5~\log d_L(z, \Omega _M,\Omega _\Lambda, \omega)
+ \frac{15}{4(1+\omega)}\log\big(1+z\big)
\end{equation}
which reduces to the standard relation as $\omega \rightarrow \infty$.
>From the last term alone we can see that $\omega \simeq 5$ can reduce 
the apparent luminosity by $\Delta m \simeq 0.2$, which is roughly
what is needed to explain the SNeIa results without a cosmological
constant. For illustrative purposes figure \ref{fig1} shows the 
above relation for two representative cosmological models, including 
the effects of $\omega$ in $d_L$, for $\omega= \pm 5$ (dotted lines) 
and the standard $(\omega=\infty)$ case (solid line).

\begin{figure}
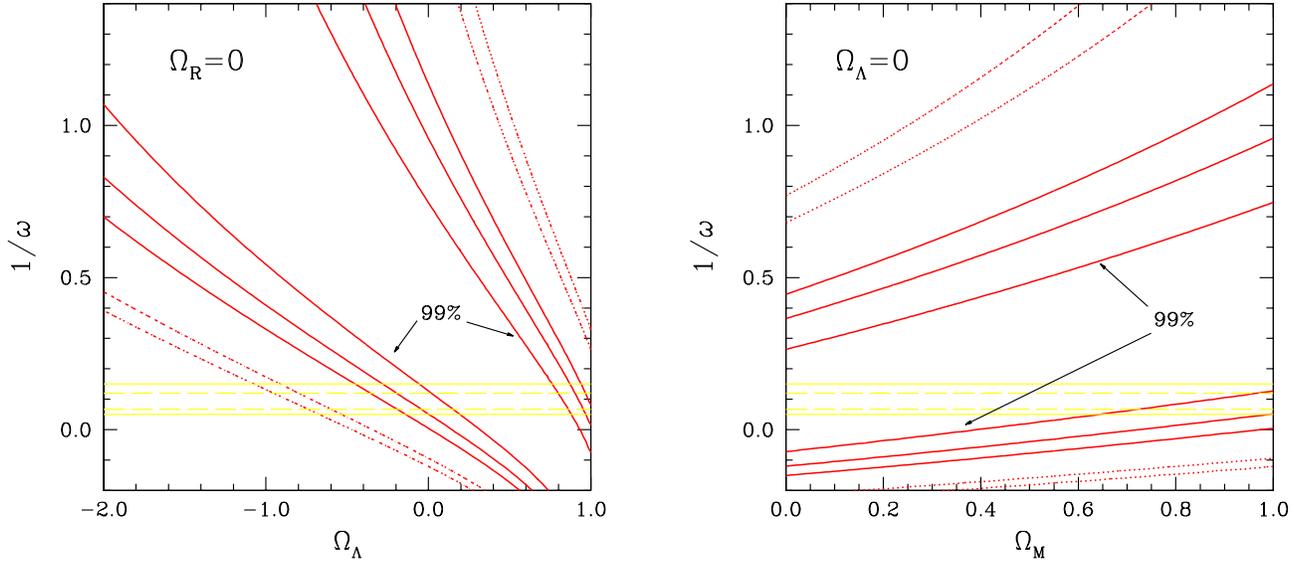

\begin{minipage}{8.5cm}
\epsfxsize 8.5cm
\epsfbox{snIa.ps2a}
\end{minipage}
\hfill
\begin{minipage}{8.5cm}
\epsfxsize 8.5cm
\epsfbox{snIa.ps2b}
\end{minipage}
\caption{Confidence contours in the plane $(\omega ,\Omega _{\Lambda})$
for a flat case $\Omega_R=0$ (left panel) and in the plane $(\omega 
,\Omega _{M})$ for the case $\Omega_\Lambda=0$ (right panel).}
\label{fig2}
\end{figure}

The effect of a varying $G$ on the time scales of SNeIa can be obtained 
from Eq.[\ref{risetime}]. Since $\tau\propto G^{-3/4}$, the ratio of
the local time scale, $\tau_0$, to the faraway one is:

\begin{equation}
\Big\langle\frac{\tau}{\tau_0}\Big\rangle\simeq
\Big\langle\frac{G}{G_0}\Big\rangle^{-3/4} = 
\langle 1+z\rangle^{-\frac{3}{4(1+\omega)}}.
\label{width}
\end{equation}
and, to make some quantitative estimates, we can use the mean 
evolution found by Riess et al. (1999a,b). From their figure 1 
we obtain the following widths of the light curve when the supernova 
is 2.5 magnitudes fainter than the peak luminosity: $\tau_0 = 45.0 
\pm 0.15$ (at $z\simeq 0$) and $\tau = 43.8 \pm 0.40$ (at $z \simeq 
0.5$), were the errors in the widths have been ascribed solely to 
the errors in the risetimes. Thus, from Eq.[\ref{width}] we obtain
$\omega \simeq 10.25^{+9.25}_{-3.65}$ ($2\sigma$ errors). Therefore,
a very small variation of the gravitational constant can account for
the reported differences in the SNeIa time scales. However these 
limits on $\omega$ should be considered as weak, in the sense that
since most SNeIa are discovered close to its peak luminosity the 
width of the light curve is poorly determined. These values are 
shown as horizontal dashed ($1\sigma$) and continuous ($2\sigma$) 
lines in Fig. \ref{fig2} where the confidence contours (at the 
99\%, 90\%, 68\% --- solid lines --- 5\% and 1\% confidence level
--- dotted lines) in the $(\omega,\Omega_{\Lambda})$ plane for a 
flat $\Omega_R=0$ universe (left panel) and in the $(\omega,\Omega 
_{M})$ plane for the case $\Omega_\Lambda=0$ (right panel) are shown.

\section{Discussion and Conclusions}

In astrophysics and cosmology the laws of physics (and in particular 
the simplest version of GR) are extrapolated outside its observational 
range of validity. It is therefore important to test for deviations of 
these laws at increasing cosmological scales and times (redshifts). 
SNeIa provide us with a new tool to test how the laws of gravity and 
cosmology were in farway galaxies ($z \simeq 0.5$). In particular, 
current limits on the (parametrized) Post Newtonian formalism 
mostly restrict to our very local Universe (see Will 1993). The 
observational limits on $\dot G/G$ come from quite different times 
and scales (see Barrow \& Parsons 1996 for a review), but mostly
in the local and nearby enviroments at $z \simeq 0$ (solar system, 
binary pulsars, white dwarf cooling, neutron stars) typical bounds 
give $\dot G/G \la 10^{-11}-10^{-12}$ yr$^{-1}$, or $\omega \ga 10-100$. 
However, STTs predict $\omega=\omega(\phi)$. That is, $\omega$ is 
not required to be a constant, so that $\omega$ can increase with 
cosmic time, $\omega=\omega(z)$, in such a way that it could approach 
the GR predictions ($\omega \rightarrow \infty$) at present time and 
still give significant deviations at earlier cosmological times. In 
this sense bounds from primordial nucleosynthesis could provide 
an important test. Current bounds on $\omega$ from nucleosynthesis
are comparable to the local values but these bounds are model 
dependent and also involve very large extrapolations.

Our analysis indicates that if we adopt the constraints derived 
from the width of the light curves of SNeIa then our best fit 
to the data requires $\omega\simeq 10$ (or equivalently $\dot{G}/G
\sim 10^{-11}$ yr$^{-1}$ or $\sim 10\%$ in $G$). 
This value is slightly smaller than 
some of the the current constraints at $z \simeq 0$, 
but it corresponds to higher redshifts $z \simeq 0.5$ and could be 
accomodated in STTs with $\omega=\omega(\phi)=\omega(z)$. If this is the case, 
at the $2\sigma$ confidence level we obtain $0.0\la\Omega_\Lambda\la 
1.0$ and the Hubble diagram of SNeIa poorly constrains $\Omega_{\rm 
M}\la 1$. At the $1\sigma$ confidence level we obtain $0.2\la
\Omega_\Lambda\la 0.8$ and $\Omega_{\rm M}\la 0.7$. If we do not 
take into account the restrictions derived from the width of the 
light curves then our conclusions are much weaker: the observational 
data and the theory can be reconciled in the framework of a 
cosmological theory with a varying $G$ with no cosmological 
constant $(\Omega_\Lambda=0)$ only if $\omega\ga 1.5$. If we 
further require a flat $\Omega_{\rm R}=0$ universe then $1.5\la
\omega\la 3.0$ is needed.

Obviously more work is needed both regarding other observational
consequences of STTs and on the physics of supernovae. In particular,
an improvement of our knowledge of the physics of thermonuclear
supernovae would provide us with an unique tool to test fundamental
laws of physics over cosmological distances. In addition it should
be stressed that new observations of distant supernovae, 
or other standard candles, at 
higher redshifts ($z>1$) could constrain even more the current limits on
the variation of the fundamental constants.

\newpage

\newpage

\end{document}